\documentclass[prb,onecolumn,superscriptaddress,amsmath,amssymb,floatfix]{revtex4-2}
\usepackage[margin=1in]{geometry}
\usepackage{amsmath}
\usepackage{xcolor}
\usepackage{graphicx}
\usepackage{siunitx}
\usepackage{listings}

\begin{document}
\title{Comment on ``Coexistence of superconductivity and topological aspects in beryllenes'', Materials Today Physics 38, 101257 (2023)}

\author{Mikhail Petrov}
\affiliation{Department of Mechanical Engineering, Tufts University, Medford, MA 02155, USA}
\author{Milorad V. Milo\v{s}evi\'c}\email{milorad.milosevic@uantwerpen.be}
\affiliation{Department of Physics, University of Antwerp, Groenenborgerlaan 171, B-2020 Antwerp, Belgium}

\begin{abstract}
In a recent publication by Li \textit{et al.} \cite{Li}, two phases of beryllene - $\alpha$ and $\beta$ - were predicted to be single-gap superconductors with critical temperatures of 9.9 K and 12.6 K respectively. Moreover, the $\alpha$-beryllene was shown to host type-I Dirac fermions with the existence of nontrivial edge states. We observe significantly weaker superconducting properties of both beryllene configurations. We argue that the superconducting gap evolution with temperature, as shown in Figure 5 (b and d) of Ref. \cite{Li}, exhibits clearly unphysical trends with increasing temperature, leading to significantly overestimated values of the critical temperature and erroneous conclusions concerning the two-gap superconducting nature of $\beta$-beryllene. On a positive note, we report the value of the gap in the Dirac cone of the topological states of interest that exceeds the temperature range of superconductivity in $\alpha$-beryllene, supporting the coexistence of topological features and superconductivity in this material.
\end{abstract}

\maketitle

\section*{Computational details}
Electronic, vibrational and structural properties of the beryllenes were calculated with density functional (perturbation) theory as implemented in Quantum Espresso (QE) version 7.2 \cite{Giannozzi_2009,Giannozzi_2017}, using the Perdew-Burke-Ernzerhof (PBE) functional
and norm-conserving pseudopotentials from the PseudoDojo project \cite{VANSETTEN201839}. Kinetic energy cutoffs for wavefunctions and for charge density and potential were set to 100 Ry and 400 Ry respectively. More than $15 \si{\angstrom}$ of vacuum was allowed in the $z$-direction to simulate 2D monolayers (independent from its periodic images in $z$-direction). We used 24 $\times$ 24 $\times$ 1 $k$-point grid and 12 $\times$ 12 $\times$ 1 $q$-point grid for electronic and vibrational properties respectively. A range of electronic smearing values - 0.02, 0.025, 0.03 and 0.04 Ry -  was considered for both beryllene phases in order to obtain vibrational dispersion without numerical instabilities. After this screening, electronic smearing of 0.04 Ry was selected optimal for $\beta$-beryllene, and of 0.03 Ry for $\alpha$-beryllene. 

Wannier90 (W90) package \cite{MOSTOFI2008685} was used to compute maximally-localized Wannier functions (MLWFs) for the beryllenes. We used 500 hundred steps for both disentanglement and minimization procedures. The initial projections were chosen to be $s$ and $p$ orbitals. The resulting spreads of the MLWFs were below $2 \si{\angstrom}^2$. Band structures obtained in W90 were ensured to match those obtained in QE. Based on the obtained Wannier functions, topological properties were determined by computing the Z$_2$ invariant via evolution of Wannier charge centers (WCC) \cite{PhysRevB.83.035108}, as implemented in the WannierTools package \cite{WU2018405}.

Superconducting properties of the beryllenes were determined by solving fully anisotropic Migdal-Eliashberg equations, as implemented in the Electron-Phonon Wannier (EPW) package version 5.7 \cite{Margine2013,Ponce2016,Giustino2007}. The screened Coulomb repulsion parameter $\mu^*$ was chosen to be 0.1, same as in Ref. \cite{Li}. Interpolated fine meshes of 200 $\times$ 200 $\times$ 1 $k$- and $q$-points were used for solving Migdal-Eliashberg equations. The width of the Fermi surface window (fsthick in EPW) was set to 0.6 eV which is around 6 times larger than the maximal vibrational frequency of the beryllenes. The cutoff for the fermion Matsubara frequencies (wscut in EPW) of 0.2 eV was used. Smearing in the energy-conserving delta functions (degaussw in EPW) was set to 0.1 eV. In order to accurately evaluate multigap nature of the superconducting state in $\beta$-beryllene, superconducting density of states (SDOS) was computed for a dense grid of energies (controlled by nqstep in EPW) containing 5000 points.

\section*{Structural, electronic and topological properties}
\begin{figure}[t]
\centering 
\includegraphics[width=0.7\linewidth]{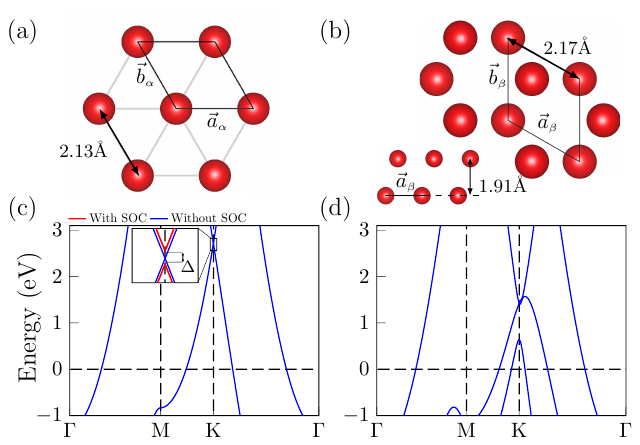}
\caption{(a,b) Crystal structures and (c,d) electronic band structures of $\alpha$ and $\beta$-beryllenes respectively. The inset of (c) shows the zoom-in on the electronic dispersion in the vicinity of the Dirac cone, computed with and without spin-orbit coupling. With spin-orbit coupling, a gap of $\Delta$ = 1 meV opens in the cone.}
\label{fig1} 
\end{figure}   

The corresponding structures of the two beryllene phases are shown in Fig. \ref{fig1}(a,b). $\alpha$-Beryllene has a hexagonal lattice with the Be-Be interatomic distance of 2.13 $\si{\angstrom}$. $\beta$-Beryllene constitutes a non-planar hexagonal lattice with two atoms per unit cell. This structure can be seen as two $\alpha$-beryllene-like layers separated by 1.91 $\si{\angstrom}$ in $z$-direction, stacked on top of each other with a small relative shift. Within each $\alpha$-like layer, Be-Be interatomic distance is 2.17 $\si{\angstrom}$. These results are in good agreement with prior reports \cite{doi:10.1021/acs.jpclett.0c02426} and Ref. \cite{Li}.

In terms of electronic properties, the two beryllene phases are rather similar. There are two types of states crossing the Fermi level [cf. Fig. \ref{fig1} (c,d)]. States of the first type are located in the vicinity of the $\Gamma$ point and are predominantly of $p_{\mathrm{z}}$ orbital character. States of the second type are located on both sides of the K points and stem from $p_{\mathrm{x,y}}$ orbitals. Similarity of the band structure close to the Fermi level is also well represented by the corresponding Fermi surfaces [shown in Fig. \ref{fig3}] which in both cases consist of a circular sheet around the $\Gamma$ point and differently shaped pockets around the K point. However, the density of states (DOS) at the Fermi level is significantly higher in $\beta$-beryllene compared to the $\alpha$ phase (0.28 and 0.18 states/eV respectively).

$\alpha$-Beryllene phase possesses a type-I Dirac band crossing located around 2.5 eV above the Fermi level [zoomed on within Fig. \ref{fig1} (c)]. By taking spin-orbit coupling (SOC) into account in the calculation, we find that a gap $\Delta$ of around 1 meV opens in the Dirac cone. Such energy corresponds to a temperature $\Delta$/$k_{\mathrm{B}}$ of roughly 12 K. To confirm the non-trivial topology, we computed the Z$_2$ topological invariant \cite{PhysRevB.83.035108}. The resulting Z$_2$ is 1 confirming the non-trivial topology of the Dirac crossing. 

\begin{table}[b]
\centering
\begin{tabular}{|p{4.2cm}|c|c|c|}
\hline
\bfseries  & $\lambda$ & Isotropic T$_c$ [K] \\ 
\hline
 $\alpha$-beryllene with 0.02 Ry & 0.53 & 4.35  \\
 \underline{$\alpha$-beryllene with 0.03 Ry} & \underline{0.46} & \underline{2.78}  \\
 $\alpha$-beryllene with 0.04 Ry & 0.43 & 2.18  \\
 $\beta$-beryllene with 0.02 Ry & 0.53 & 6.15  \\
 \underline{$\beta$-beryllene with 0.04 Ry} & \underline{0.52} & \underline{6.09}  \\ [0.1cm]
\hline
\end{tabular}
\caption{Comparison of the total electron-phonon coupling $\lambda$ and the isotropic critical temperatures for the two beryllene phases (using the Allen-Dynes formula), computed with different values of the electronic smearing set in QE for electronic and vibrational properties calculations. The underlined cases are selected for further anisotropic Migdal-Eliashberg calculations of superconductivity.
}
\label{tab1}
\end{table}

\section*{Phonon dispersions, electron-phonon coupling and superconducting properties}
\begin{figure}[t]
\centering 
\includegraphics[width=0.6\linewidth]{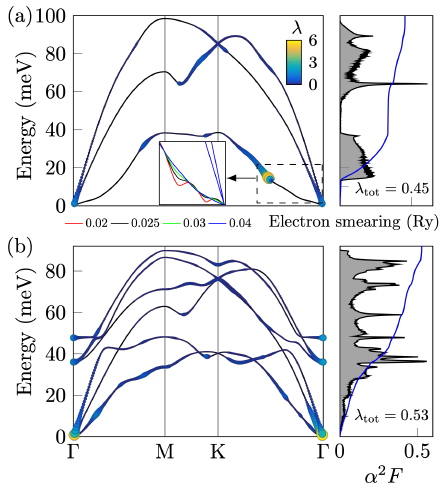}
\caption{Phonon dispersion with mode-resolved electron-phonon coupling (EPC) $\lambda_{\mathbf{q}\nu}$ indicated by both color and dot size (left panels), and the isotropic Eliashberg function $\alpha^2F$ with isotropic EPC function $\lambda$ (right panel) for (a) $\alpha$-beryllene and (b) $\beta$-beryllene. The colorbar as shown in (a) serves for both beryllene phases. The inset in (a) reflects the effect of electronic smearing on phonon dispersion in the region marked by dashed rectangle. }
\label{fig2} 
\end{figure}

Both beryllene phases are dynamically stable as can be seen from the absence of negative frequencies in their phonon dispersions [Fig.~\ref{fig2} (a,b)]. However, we found that the dispersion of the lowest acoustic modes in the vicinity of the $\Gamma$ point is sensitive to the electronic smearing (degauss parameter in QE, not to be confused with the smearing variables implemented in EPW). To investigate the effect that the electronic smearing value has on the predicted superconducting properties, we performed a complete sequence consisting of structure optimization, electronic/phonon and EPW calculations for each smearing value we chose. For $\alpha$-beryllene, smaller values of electron smearing result in an oscillatory behavior of the lowest acoustic phonon mode [as shown in inset of Fig.~\ref{fig2}(a)]. As the value of the smearing is increased, phonon dispersion becomes more stable and the smearing of 0.03 Ry is set as optimal for the $\alpha$ configuration. For $\beta$-beryllene, smaller values of the smearing result in small negative energies in the lowest acoustic mode around the $\Gamma$ point, which persist for the values of smearing lower than 0.04 Ry. Thus, the smearing value of the 0.04 Ry is set as optimal for the $\beta$ configuration. Phonon band structures shown in Fig. \ref{fig2} (a,b) were computed with the smearing values as defined above. While instabilities do not have a profound impact on the EPC in $\beta$-beryllene, they do increase it significantly in $\alpha$-beryllene (nearly twice, cf. Table \ref{tab1}), which is the case because the smearing-sensitive lowest acoustic mode exhibits one of the most profound contributions to the total EPC.

\begin{figure}[t]
\centering 
\includegraphics[width=0.9\linewidth]{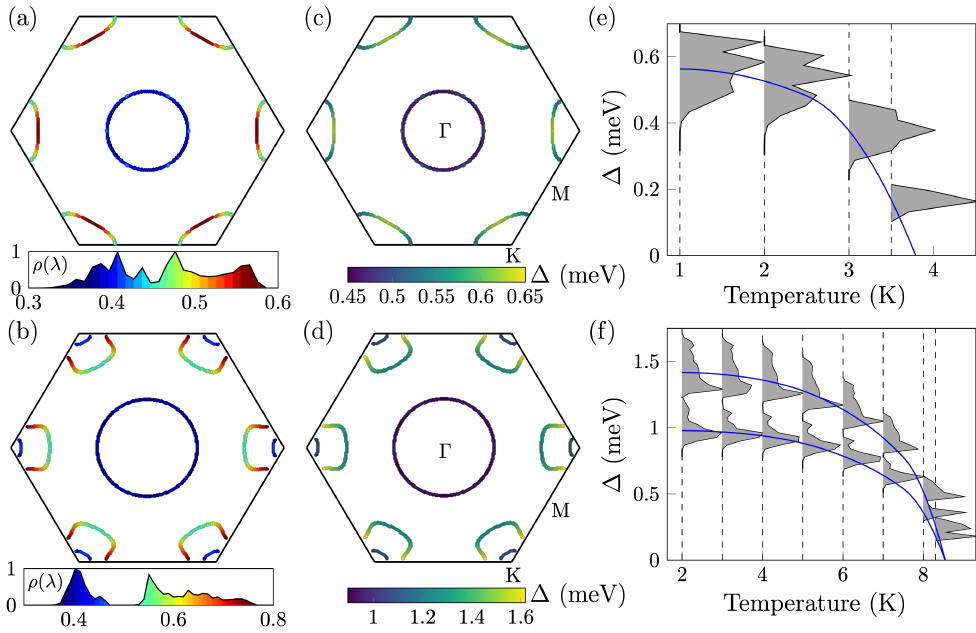}
\caption{Electron-phonon coupling (EPC) plotted on the Fermi surface and the normalized distribution of EPC $\rho(\lambda)$, for (a) $\alpha$-beryllene and (b) $\beta$-beryllene. The $\rho(\lambda)$ plot also serves as the colorbar for the EPC on the Fermi surface. (c,d) The superconducting gap plotted on the Fermi surface, for $\alpha$-beryllene at 1 K and $\beta$-beryllene at 2 K respectively.
(e,f) Anisotropic superconducting gap spectrum as a function of temperature, for $\alpha$-beryllene and $\beta$-beryllene respectively. The weighted averages of different domes are traced with solid lines.}
\label{fig3} 
\end{figure}

The calculated values of EPC are not particularly large, but the coupling is sufficiently strong for both beryllene configurations to be conclusively superconducting. To map out the origins of superconductivity from the available electronic states, we consistently solved the fully anisotropic Migdal-Eliashberg equations.
In Fig. \ref{fig3} (a)-(d) we show the distribution of the EPC and the superconducting gap plotted on the Fermi surface, for both studied phases of beryllene. 

In order to find the critical temperature, we solved the anisotropic Migdal-Eliashberg equations for a range of temperatures, where the critical temperature is determined as the lowest temperature for which there were no solutions of the equations for the superconducting gap. As shown in Fig. \ref{fig3} (e), $\alpha$ - beryllene is a single-gap superconductor with the critical temperature around 3.8 K, which is slightly above the predicted T$_{\mathrm{c}}$ of 2.78 K from the isotropic Allen-Dynes formula (given in Table \ref{tab1}). The superconducting gap spectrum is clearly anisotropic and spread over a range of energies. In particular, $p_{\mathrm{z}}$ states located around the $\Gamma$ point show smaller values of the superconducting gap $\Delta$ compared to the $p_{\mathrm{x,y}}$ states.

Even though $\beta$-beryllene does not exhibit significantly stronger EPC, its critical temperature obtained from the anisotropic Migdal
-Eliashberg calculations reaches significantly higher - around 8.5 K [Fig. \ref{fig3} (f)]. However, it is also only slightly higher than the predicted 6 K in the isotropic Allen-Dynes limit. Higher critical temperature of the $\beta$ phase can be at least partly attributed to the higher electronic density of states at the Fermi level, which is about 50\% larger than that of the $\alpha$-beryllene. The superconducting gap distribution is highly anisotropic, with two distinct domes. The lower dome stems from the $p_{\mathrm{z}}$ states located around the $\Gamma$ point and a smaller Fermi sheet located around K points whereas the higher dome stems predominantly form the larger Fermi sheet around K points. The resulting gap spectrum distribution suggests two separate gaps, but the separation between the domes associated with those gaps is small and effectively vanishes at higher temperatures [cf. Fig. \ref{fig3} (f)]. Nevertheless, by carefully evaluating superconducting density of states (SDOS) on a dense grid of energies (see Computational Details), we confirm the two-gap nature of the superconducting state of $\beta$-beryllene [see Fig. \ref{fig4}]. The SDOS results presented in Figure S5 of the supplementary material to Ref. \cite{Li} erroneously suggest the single-gap behavior, likely due to the lack of resolution at the energies where two domes in the gap spectrum are located.

\begin{figure}[t]
\centering 
\includegraphics[width=0.6\linewidth]{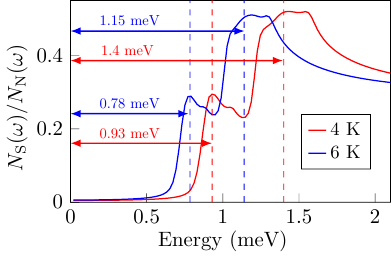}
\caption{Quasiparticle density of states in the superconducting state (SDOS), relative to the DOS in the normal state $N_{\textrm{S}}$($\omega$)/$N_{\textrm{N}}$($\omega$) for $\beta$-beryllene, computed at 4 K and 6 K. Dashed lines mark the weighted average values of different domes in the anisotropic superconducting gap spectrum at the corresponding temperatures [cf. Fig. \ref{fig3} (f)].}
\label{fig4} 
\end{figure} 

\section*{Summary}
Our results concerning the structural and electronic properties of beryllenes are in good agreement with Li \textit{et al.} \cite{Li}. An actual opening of the gap in the Dirac cone upon taking into account spin-orbit coupling was not shown in Ref. \cite{Li}, neither was the gap energy value indicated. We point out that the gap exceeds 1 meV and suggests its robustness in the temperature range where superconductivity is expected. We agree with Ref. \cite{Li} on the Z$_2$ invariant of that Dirac crossing, thus confirming the existence of the non-trivial states there. 

For vibrational properties, we observed high sensitivity of the lowest acoustic phonon mode dispersion in the close proximity of the $\Gamma$ point to the electronic smearing. In order to obtain convincingly stable phonon dispersion, we had to increase the smearing values to 0.03 and 0.04 Ry for $\alpha$ and $\beta$ beryllenes respectively. Ref. \cite{Li} reports stable phonons already for lower 0.02 Ry smearing.

Despite the differences in vibrational dispersion, we observed near perfect agreement of the EPC in our calculations and in Ref. \cite{Li}. Moreover, taking the values of $w_{\textrm{log}}$ and $\lambda$ obtained in EPW by Li \textit{et al.} (Table 2 of Ref. \cite{Li}) and substituting them into the Allen-Dynes formula, yielded excellent agreement between our results and Ref. \cite{Li} for the isotropic critical temperature. 

However, the most striking difference arose in the anisotropic calculations. As expected, we found a monotonic decrease of the superconducting gap energy with temperature for both beryllene phases, with the critical temperature roughly corresponding the BCS relation to the gap at zero temperature. Surprisingly, Ref. \cite{Li} shows an initial upward trend of $\Delta(T)$, thus unphysically pushing the critical temperature to much higher values. Additionally, for $\alpha$-beryllene, at the lowest considered temperature of 1 K, we observed the superconducting gap spectrum reaching at most 0.7 meV (cf. Fig. \ref{fig3} (e)). In contrast, Li \textit{et al.} showed the gap values reaching above 0.9 meV for 2 K even before the onset of the upward trend in $\Delta(T)$ (Figure 5(b) of Ref. \cite{Li}). As a consequence, the Migdal-Eliashberg anisotropic critical temperature for $\alpha$-beryllene in our work is only about 40\% of the value reported in Ref. \cite{Li} (3.8K vs. 9.9K). The same arguments can be applied to the $\beta$-beryllene, where our Migdal-Eliashberg result for the critical temperature is about 65\% of the value reported in Ref. \cite{Li} (8.5K vs. 12.6K). Lastly, we dispute the claimed single-gap behavior of $\beta$-beryllene in Ref. \cite{Li}. $\beta$-Beryllene is a two-gap superconductor, which is confirmed by careful evaluation of SDOS in our work.

\section*{NOTE ON SCIENTIFIC ETHICAL CONDUCT}
This comment was sent to authors and the journal in question in an effort to make a correction in the public record, without any malicious intent. The authors did not opt to respond in any way or make the correction accordingly, and the journal does not publish corrections or comments. The Editorial Board of the journal was in contact with the authors and has decided that correction is not warranted.

\section*{Acknowledgments}
This work was supported by the Research Foundation-Flanders (FWO). The computational resources and services were provided by the VSC (Flemish Supercomputer Center), funded by the FWO and the Flemish Government -- department EWI.

\bibliographystyle{unsrt}
\bibliography{biblio}

\end{document}